\documentclass{aastex6}
\begin{document}
\title{External Shock in a Multi-Bursting Gamma-ray Burst: Energy Injection Phase induced by the Later Launched Ejecta}
\author{Da-Bin Lin\altaffilmark{1,2}, Bao-Quan Huang\altaffilmark{1,2}, Tong Liu\altaffilmark{3}, Wei-Min Gu\altaffilmark{3}, Hui-Jun Mu\altaffilmark{3}, and En-Wei Liang\altaffilmark{1,2}}
\altaffiltext{1}{GXU-NAOC Center for Astrophysics and Space Sciences, Department of Physics, Guangxi University, Nanning 530004, China; lindabin@gxu.edu.cn}
\altaffiltext{2}{Guangxi Key Laboratory for the Relativistic Astrophysics, Nanning 530004, China}
\altaffiltext{3}{Department of Astronomy, Xiamen University, Xiamen, Fujian 361005, China}
\begin{abstract}
Central engine of gamma-ray bursts (GRBs) may be intermittent
and launch several episodes of ejecta separated by a long quiescent interval.
In this scenario, an external shock is formed due to the propagation of the first launched ejecta into the circum-burst medium
and the later launched ejecta may interact with the external shock at later period.
Owing to the internal dissipation, the later launched ejecta may be observed at a later time ($t_{\rm{jet}}$).
In this paper, we study the relation of $t_{\rm{b}}$ and $t_{\rm{jet}}$,
where $t_{\rm{b}}$ is the collision time of the later launched ejecta with the formed external shock.
It is found that the relation of $t_{\rm{b}}$ and $t_{\rm{jet}}$ depends on
the bulk Lorentz factor ($\Gamma_{\rm{jet}}$) of the later launched ejecta
and the density ($\rho$) of the circum-burst medium.
If the value of $\Gamma_{\rm{jet}}$ or $\rho$ is low,
the $t_{\rm{b}}$ would be significantly larger than $t_{\rm{jet}}$.
However, the $t_{\rm{b}}\sim t_{\rm{jet}}$ can be found
if the value of $\Gamma_{\rm{jet}}$ or $\rho$ is significantly large.
Our results can explain the large lag of the optical emission relative to the
$\gamma$-ray/X-ray emission in GRBs, e.g., GRB~111209A.
For GRBs with a precursor, our results suggest that
the energy injection into the external shock and thus
more than one external-reverse shock may appear in the main prompt emission phase.
According to our model, we estimate the Lorentz factor of the second launched ejecta in GRB~160625B.
\end{abstract}
\section{Introduction} \label{sec:intro}
Observationally, gamma-ray bursts (GRBs) generally appear as a powerful burst of $\gamma$-rays
followed by a long-lived afterglow emission.
The light curves of afterglow emission usually can be decomposed into
four power-law segments, i.e.,
an initial steep decay, a shallow decay, a normal decay, and a late steeper decay,
together with one or several flares (\citealp{Zhang_B-2006-Fan_YZ}; \citealp{Nousek_JA-2006-Kouveliotou_C}; \citealp{O'Brien_PT-2006-Willingale_R}; \citealp{Zhang_BB-2007-Liang_EW}).
Theoretically, the phenomena of GRBs can be understood as follows.
The central engine of a GRB, such as a stellar-mass black hole surrounded by a hyper-accretion disc
(e.g., \citealp{Narayan_R-1992-Paczynski_B}; \citealp{Popham_R-1999-Woosley_SE}; \citealp{Narayan_R-2001-Piran_T};
\citealp{Gu_WM-2006-Liu_T}; \citealp{Liu_T-2007-Gu_WM})
or a millisecond magnetar (\citealp{Usov_VV-1992}; \citealp{Thompson_C-1994}; \citealp{Dai_ZG-1998a-Lu_T};
\citealp{Wheeler_JC-2000-Yi_I}; \citealp{Zhang_B-2001a-Meszaros_P};
\citealp{Metzger_BD-2008-Quataert_E}; \citealp{Metzger_BD-2011-Giannios_D};
\citealp{Bucciantini_N-2012-Metzger_BD}; \citealp{Lu_HJ-2014-Zhang_B}; \citealp{Mosta_P-2015-Ott_CD}),
launches a relativistical ejecta,
which may be composed of many mini-shells with different Lorentz factors.
Then, the internal shocks (\citealp{Rees_MJ-1994-Meszaros_P}) or internal-collision-induced magnetic reconnection and turbulence
(\citealp{Zhang_B-2011-Yan_H}; \citealp{Deng_W-2015-Li_H})
can be formed due to the collisions of shells.
Owing to the powerful collisions,
the prompt $\gamma$-rays are produced.
The observed prompt $\gamma$-rays may also be released near the photosphere,
where the ejecta becomes transparent for thermal photons.
When the relativistic ejecta further propagates into the circum-burst medium,
an external shock would be developed and thus
produces a long-term broadband afterglow emission
(\citealp{Sari_R-1998-Piran_T}; \citealp{Meszaros_P-1999-Rees_MJ}; \citealp{Sari_R-1999a-Piran_T,Sari_R-1999b-Piran_T}).
If there is no energy injection into the external shock,
a normal decay would appear.
However,
the decay may become shallow during the continuous energy injection into the external shock
(\citealp{Zhang_B-2006-Fan_YZ}; \citealp{Nousek_JA-2006-Kouveliotou_C}; \citealp{Panaitescu_A-2006-Meszaros_P}).
This is the origin of the normal decay and shallow decay observed in the canonical light curve of afterglow emission.
The initial steep decay phase in the afterglows is believed to be the tail emission of the prompt $\gamma$-rays (\citealp{Barthelmy_SD-2005-Cannizzo_JK}; \citealp{Liang_EW-2006-Zhang_B}; \citealp{O'Brien_PT-2006-Willingale_R}),
and the late shallower decay is the external shock emission after the jet break phase.

The X-ray flares generally show sharp rise with a steep decay and thus may not be produced in the external shock.
Similar to the formation of the prompt $\gamma$-rays,
most of the X-ray flares are believed to be the internal origin
(e.g., \citealp{Romano_P-2006-Moretti_A}; \citealp{Falcone_AD-2006-Burrows_DN};
\citealp{Burrows_DN-2005-Romano_P}; \citealp{Falcone_AD-2006-Burrows_DN,Falcone_AD-2007-Morris_D};
\citealp{Zhang_B-2006-Fan_YZ}; \citealp{Nousek_JA-2006-Kouveliotou_C}; \citealp{Liang_EW-2006-Zhang_B};
\citealp{Chincarini_G-2007-Moretti_A,Chincarini_G-2010-Mao_J}; \citealp{Hou_SJ-2014-Geng_JJ}; \citealp{Wu_XF-2013-Hou_SJ}; \citealp{Yi_SX-2015-Wu_XF}; \citealp{Yi_SX-2016-Xi_SQ}; \citealp{Mu_HJ-2016b-Lin_DB}; \citealp{Mu_HJ-2016a-Gu_WM}).
It should be noted that the external-reverse shock (RS) is also adopted to explain
the X-ray flares or prompt $\gamma$-rays in some burst (e.g., \citealp{Shao_L-2005-Dai_ZG,Kobayashi_S-2007-Zhang_B,Fraija_N-2015,Fraija_N-2016-Lee_WH}).
Besides the X-ray flares,
the X-ray plateau, e.g. GRBs~070110 and 060202 (\citealp{Troja_E-2007-Cusumano_G}; \citealp{Liang_EW-2007-Zhang_BB}; \citealp{Lu_HJ-2014-Zhang_B}), or
X-ray bump, e.g. GRBs~121027A and 111209A (\citealp{Wu_XF-2013-Hou_SJ}; \citealp{Stratta_G-2013-Gendre_B}; \citealp{Yu_YB-2015-Wu_XF}),
may also have the same physical origin as the prompt $\gamma$-rays.
Thus, these X-ray plateau/bump are always dubbed ``internal plateau/bump''.
Most commonly, GRBs have a single episode of prompt $\gamma$-rays.
However, some bursts show two or three episodes of prompt $\gamma$-rays
separated by a long quiescent interval ($\sim 100$~s), such as GRBs~110709B (\citealp{Zhang_BB-2012-Burrows_DN}) and 160625B (\citealp{Zhang_BB-2016-Zhang_B}; \citealp{Lu_HJ-2017-Lu_J}; \citealp{Alexander_KD-2017-Laskar_T}; \citealp{Fraija_N-2017-Veres_P}).
In addition, 10\% GRBs have a precursor emission,
which may have the same physical origin as the prompt $\gamma$-rays (\citealp{Troja_E-2010-Rosswog_S, Hu_YD-2014-Liang_EW}).
These observations suggested that the central engine of GRBs may be intermittent
and launch several episodes of ejecta separated by a long quiescent interval.
In this case, an external shock would be formed
during the propagation of the first launched ejecta (JET1) into the circum-burst medium.
The remnants of the later launched ejecta would catch up with the formed external shock at later period
and thus the energy injection into the external shock would appear.
The radiative signature associated with this process would appear
through an external-forward shock
(\citealp{Sari_R-1998-Piran_T, Sari_R-1999b-Piran_T}) and/or an RS
(\citealp{Meszaros_P-1999-Rees_MJ}; \citealp{Sari_R-1999a-Piran_T,Sari_R-1999b-Piran_T}).
In this work, we focus on the observed time of the energy injection into the external shock.

The paper is organized as follows.
In Section~\ref{Sec:Dynamics}, we summarize the equations governing the evolution of the external shock.
Here, the evolution of the external shock is estimated with energy injection from
the remnants of the second launched ejecta (JET2).
In Section~\ref{Sec:results}, we estimate the observed time of the energy injection into the formed external shock.
In Section~\ref{Sec:Conclusion_and_Discussion},
the conclusions and discussions are presented.

\section{Evolution of the External Shock with Energy Injection}\label{Sec:Dynamics}
We study the evolution of the external shock in the situation that
the central engine of GRBs launches two episodes of ejecta separated by a long quiescent interval ($t_{\rm jet}$).
The external shock is formed during the propagation of JET1 into the circum-burst medium.
Then, the remnants of JET2 would catch up with the formed external shock at
the observer time $t_{\rm b}$ and thus the energy injection into the external shock would appear at $t_{\rm obs}\geqslant t_{\rm b}$.
In this section, we present the equations governing the evolution of the external shock.

Based on the conservation of energy and momentum,
one can have (\citealp{Piran_T-1999})
\begin{equation}\label{Eq:Gamma_Initial}
M'd\Gamma  =  - ({\Gamma ^2} - 1)dm
\end{equation}
and
\begin{equation}\label{Eq:Thermal_energy}
dU'=(1-\varepsilon)(\Gamma-1)dmc^2,
\end{equation}
where $\Gamma$ is the bulk Lorentz factor of the external shock,
and $M'=M'_{\rm ej}+m+U'/c^2$ is the total mass including
the initial mass $M'_{\rm ej}$ of the ejecta,
the sweep-up mass $m$ from the circum-burst,
and the internal energy $U'$ of the external shock.
Here, the physical quantities in the shell's/observer's frame are denoted with/without a prime.
The radiated thermal energy in the shell's frame is described as $\varepsilon(\Gamma-1)dmc^2$
with $\varepsilon$ being the radiation efficiency of the external shock.
The value of $\varepsilon$ is assumed as a constant during the evolution of the external shock.
Equation~(\ref{Eq:Gamma_Initial}) can also be derived based on the relation between
the total kinetic energy of the decelerated ejecta $E_{\rm k}$ and the radiated thermal energy (\citealp{Huang_YF-1999-Dai_ZG}), i.e.,
\begin{equation}\label{Eq:Total kinetic energy}
dE_{\rm k} =-\varepsilon \Gamma(\Gamma-1)dmc^2,
\end{equation}
where $E_{\rm k}=(\Gamma-1)(M'_{\rm ej}+m)c^2+\Gamma U'$.
With an energy injection $dE_{\rm inj}$, Equation~(\ref{Eq:Total kinetic energy}) can be modified as
\begin{equation}\label{Eq:Modified}
dE_{\rm k}=-\varepsilon \Gamma(\Gamma-1)dmc^2+dE_{\rm inj}.
\end{equation}
According to Equations~(\ref{Eq:Thermal_energy}) and (\ref{Eq:Modified}), one can have
\begin{equation}
M'c^2d\Gamma  = dE_{\rm{inj}} - ({\Gamma ^2} - 1)dmc^2
\end{equation}
or
\begin{equation}\label{Eq:Gamma}
\frac{{d\Gamma }}{{d{t_{{\rm{obs}}}}}} = \frac{1}{M'}\left[ {\frac{1}{{{c^2}}}\frac{{d{E_{{\rm{inj}}}}}}{{d{t_{{\rm{obs}}}}}} - ({\Gamma ^2} - 1)\frac{{dm}}{{d{t_{{\rm{obs}}}}}}} \right],
\end{equation}
which describes the evolution of $\Gamma$ with respect to the observer time $t_{\rm obs}$.
The evolution of other parameters are described as
\begin{equation}
\frac{{dU'}}{{d{t_{{\mathop{\rm obs}\nolimits} }}}} =(1-\varepsilon)(\Gamma  - 1){c^2}\frac{{dm}}{{d{t_{{\mathop{\rm obs}\nolimits} }}}},
\end{equation}
\begin{equation}\label{Eq:mass}
\frac{{dm}}{dt_{\rm obs}} = \frac{{c\beta }}{{1 - \beta }}2\pi \rho{R^2}\left( {1 - \cos {\theta _{\rm jet}}} \right),
\end{equation}
\begin{equation}\label{Eq:R}
\frac{{dR}}{{d{t_{{\rm{obs}}}}}} = \frac{{c\beta }}{{1 - \beta }},
\end{equation}
where $\beta=\sqrt{1-1/\Gamma^2}$ is the velocity of the external shock,
$\rho$ is the density of the circum-burst environment,
and $\theta_{\rm jet}$ is the half opening angle of the ejecta.
The evolution of $\theta_{\rm jet}$ is not considered in this work.
Two cases of circum-burst medium, i.e., ISM and wind, are studied.
Then, we take (e.g., \citealp{Chevalier_RA-2000-Li_ZY})
\begin{equation}
\rho=
\left\{ {\begin{array}{*{20}{c}}
{5 \times 10^{11}A_*R^{-2}\,\rm g\cdot cm^{ - 1},}&{\rm{wind},}\\
{n_0m_p\,\rm cm^{ - 3},}&{\rm{ISM},}
\end{array}} \right.
\end{equation}
with $m_p$ being the proton mass.

Due to the internal dissipation of the ejecta,
JET2 may be observed at the observer time $t_{\rm jet}$.
After this phase, the remnants of JET2 moves on and may collide with the formed external shock at later period.
Then, the energy injection into the external shock may appear.
To simplify our study, the function form of ${dE_{\rm inj}}/{dt_{\rm obs}}$ is described as
\begin{equation}\label{Eq:Energy_Injection}
\frac{{d{E_{{\rm{inj}}}}}}{{dt}} = \left\{ {\begin{array}{*{20}{c}}
{E_{\rm k,jet2}/T_{\rm inj},\;}&{t_{\rm b}< t_{\rm obs} < t_{\rm b} + T_{\rm inj},}\\
{0,}&{{\rm{others}},}
\end{array}} \right.\end{equation}
where $E_{\rm k,jet2}=10^{53}$~erg is taken as the remanent kinetic energy of JET2,
$T_{\rm inj}=\max(T_{90}, R_{\rm b}/2\Gamma_{\rm jet2}^2)$ is adopted,
$\Gamma_{\rm jet2}$ ($T_{90}=10$~s) is the Lorentz factor (duration of the prompt emission) of JET2,
and
the energy injection into the external shock is assumed to begin at the radius $R_{\rm b}$ and the observer time $t_{\rm b}$.
It should be noted that
we are interested in the value of $t_{\rm b}$ rather than the details of energy injection.

To estimate the value of $t_{\rm b}$,
we trace the locations of JET2 and the external shock at every moment.
If these two locations are the same, JET2 hits the external shock.
The corresponding observer time is the value of $t_{\rm b}$.
Then, the energy injection as Equation~(\ref{Eq:Energy_Injection}) is added into the external shock.
In our work, the evolution of the external shock is estimated with Equations~(\ref{Eq:Gamma})-(\ref{Eq:Energy_Injection})
from the radius $R_0=10^{14}$~cm.
In addition, the initial kinetic energy $E_{\rm k,0}=10^{53}\rm erg$, the initial Lorentz factor $\Gamma_0=300$,
$M'_{\rm ej}=E_{\rm k,0}/(\Gamma_0-1)$, and the redshift $z=1$ of the burst are adopted.

\section{Results}\label{Sec:results}
Figure~\ref{Fig:fig1} shows the evolution of the external shock in an ISM environment,
where the JET2 with $(t_{\rm jet}, \Gamma_{\rm jet2})=(30{\rm s}, 300)$, $(100{\rm s}, 300)$, $(30{\rm s}, 100)$, and $(100{\rm s}, 100)$ are adopted in the upper-left, upper-right, lower-left, and lower-right panels, respectively.
In this figure, the red, black, and blue solid lines represent the X-ray (0.3-10~keV) flux
from the external-forward shock in the situations with $n_0=10^{-2}$, $1$, and $10^2$, respectively.
The observed time (i.e., $[t_{\rm jet}, t_{\rm jet}+T_{90}]$)
of JET2 emission is shown with a green horizontal line,
and
the values of $t_b$
are indicated with a red, black, and blue vertical dashed lines
for situations with $n_0=10^{-2}$, $1$, and $10^2$, respectively.
From Figure~\ref{Fig:fig1}, it can be found that the observed time of the energy injection is larger than
the observed time of JET2, i.e., $t_{\rm b}>t_{\rm jet}$.
Moreover, the lower value of $\Gamma_{\rm jet2}$ or $\rho$ is,
the higher value of $t_b$ would be.
Then, we plot the relation of $t_b$ and $t_{\rm jet}$ in Figure~\ref{Fig:fig2},
where the value of $\Gamma_{\rm jet2}=300$, $100$, and $30$ are adopted in the upper, middle, and lower sub-figures, respectively.
In this figure, ISM (wind) environment is adopted in the left (right) panels,
the green dashed lines describe the relation of $t_b=t_{\rm jet}$,
and the blue, black, and red lines represent the situations with
$n_0$ (or $A_*$)$=10^{-2}$, $1$, and $10^{2}$, respectively.
According to Figure~\ref{Fig:fig2},
the value of $t_{\rm b}$ can be significantly larger than that of $t_{\rm jet}$.
Lower value of $\Gamma_{\rm jet2}$ or $\rho$ or $t_{\rm jet}$ is, the higher value of $t_{b}/t_{\rm jet}$ would be.
It is interesting to point out that $t_{\rm b}\sim t_{\rm jet}$ can be obtained if the value of $\Gamma_{\rm jet2}$ or $\rho$
or $t_{\rm jet}$ is significantly large.

The dependence of $t_b$ on $\Gamma_{\rm jet2}$, $\rho$, and $t_{\rm jet}$ can be estimated as follows.
After the internal dissipation, the JET2 is coasting forward.
The observer time corresponding to the JET2 arriving at $R$
can be estimated as
\begin{equation}
t_{\rm obs, jet2}(R)=\frac{(R - R_{\rm dis})(1 - \beta_{\rm jet2})}{c\beta_{\rm jet2}}(1+z) + t_{\rm jet},
\end{equation}
where $R_{\rm dis}=10^{14}$~cm (the dissipation location of JET2) and $\beta_{\rm jet2}=\sqrt{1-1/\Gamma_{\rm jet2}^2}$ are adopted in this paper.
With Equations~(\ref{Eq:Gamma_Initial}) and (\ref{Eq:Thermal_energy}),
the observed time of the external shock locating at $R$ can be estimated as
\begin{equation}
t_{\rm obs}(R)\approx \frac{R}{2\Gamma _0^2c{\left(1+ \beta{m\Gamma _0}/M'_{\rm ej}  \right)}^\alpha}(1+z),
\end{equation}
where $\alpha=2/(\varepsilon  - 2)$, $\beta=-(2/\alpha)[-\alpha(3-s)+1]^{1/\alpha}$, and $\rho\propto R^{-s}$ are adopted.
Then, the collision location $R_{\rm b}$ for JET2 catching up with the external shock
can be found by taking $t_{\rm obs, jet2}=t_{\rm obs}$ or
\begin{equation}\label{Eq:jet_Observed_Time}
\frac{(R_{\rm b} - R_{\rm dis})(1 - \beta_{\rm jet2})}{c\beta_{\rm jet2}}+ \frac{t_{\rm jet}}{1+z}
=
\frac{R_{\rm b}}{2\Gamma _0^2c{\left(1+ \beta{m\Gamma _0}/M'_{\rm ej}  \right)}^\alpha}.
\end{equation}
Accordingly, the observed time $t_{\rm b}$ of the collision can be estimated with
\begin{equation}\label{Eq:t_b_t_s}
t_{\rm b} = \frac{(R_{\rm b} - R_{\rm dis})(1 - \beta_{\rm jet2})}{c\beta_{\rm jet2}}(1+z)  + t_{\rm jet}.
\end{equation}
With Equations~(\ref{Eq:jet_Observed_Time})-(\ref{Eq:t_b_t_s}),
we estimate the relation of $t_b$ and $t_{\rm jet}$ for situations with $s=0$ and $\Gamma_{\rm jet2}=100$.
The results are shown with dashed lines in the middle-left panel of Figure~\ref{Fig:fig2},
where the blue, black, and red dashed lines represent the situations with
$n_0=10^{-2}$, $1$, and $10^{2}$, respectively.
It can be found that the value of $t_{\rm b}$ estimated based on Equations~(\ref{Eq:jet_Observed_Time})-(\ref{Eq:t_b_t_s})
is consistent with those estimated with Equations~(\ref{Eq:Gamma})-(\ref{Eq:R}).
According to Equation~(\ref{Eq:jet_Observed_Time}),
one can find a low $R_{\rm b}$ and thus a low $t_{\rm b}$ in the situations with a larger $\Gamma_{\rm jet2}$ or $\rho$.

\section{CONCLUSIONS AND DISCUSSIONS}\label{Sec:Conclusion_and_Discussion}
Observations reveal that the central engine of GRBs may be intermittent
and launch several episode of ejecta separated by a long quiescent interval.
Owing to the internal dissipation,
an episode of ejecta may be observed as a precursor, an episode of prompt $\gamma$-rays,
an X-ray flare, an X-ray plateau, or an X-ray bump.
In addition, an external shock is formed due to the propagation of the first launched ejecta into the circum-burst medium.
Then, the later launched ejecta would collide with the formed external shock at a later time $t_{\rm b}$.
In this paper, we study the relation of $t_{\rm b}$ and $t_{\rm jet}$,
where $t_{\rm jet}$ is the observed time of the jet emission formed in the internal dissipation processes of the later launched ejecta.
We find that the value of $t_{\rm b}$ can be significantly larger
than that of $t_{\rm jet}$.
If the bulk Lorentz factor ($\Gamma_{\rm jet2}$) of the later launched ejecta
or the density ($\rho$) of the circum-burst medium
is significantly low,
the value of $t_{\rm b}$ may be significantly larger than $t_{\rm jet}$.
However, the situation of $t_{\rm b}\sim t_{\rm jet}$ can be found
if the value of $\Gamma_{\rm jet2}$ or $\rho$ or $t_{\rm jet}$ is significantly large.
These results can explain the large lag of the optical emission relative to the $\gamma$-rays/X-rays
observed in GRBs, e.g., the X-ray flare observed at $t_{\rm obs}\sim 10^3$~s after the burst trigger in GRB~111209A.

\begin{figure}\label{fig:general}
\plotone{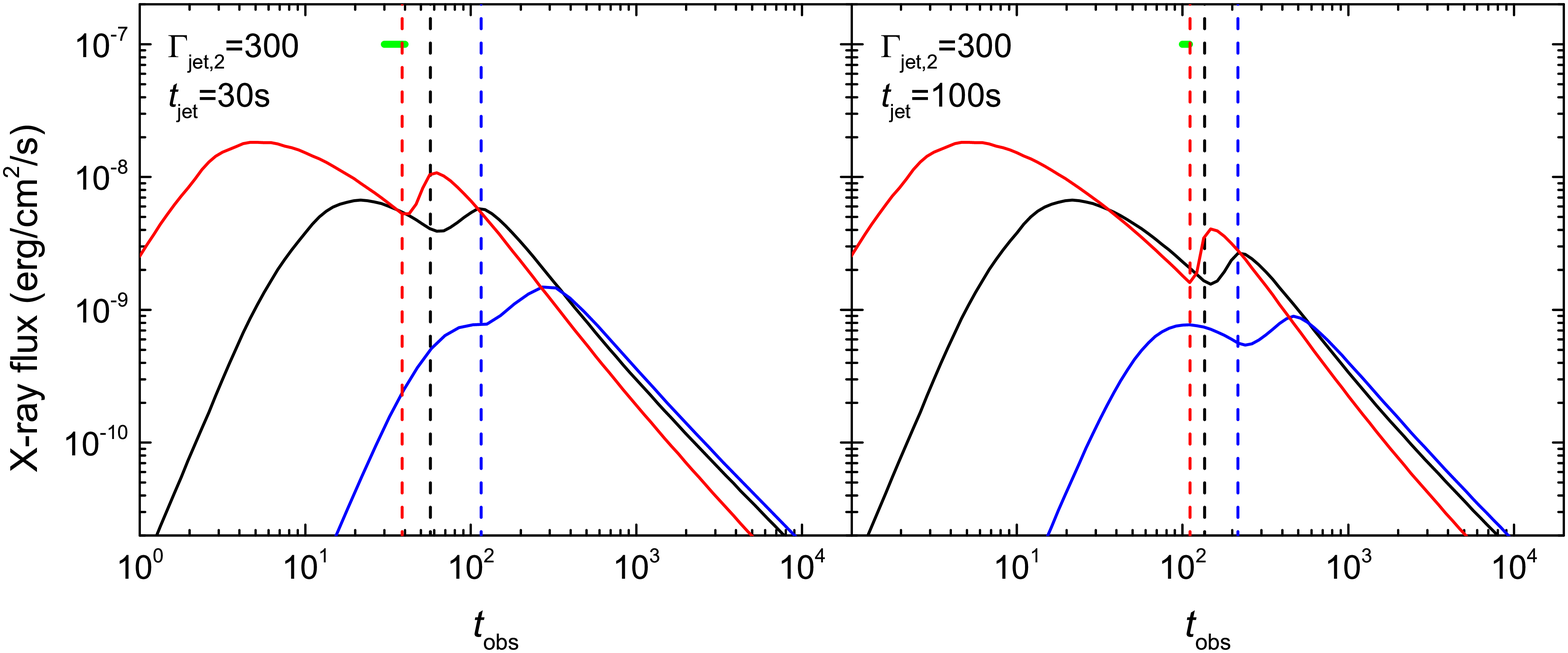}
\plotone{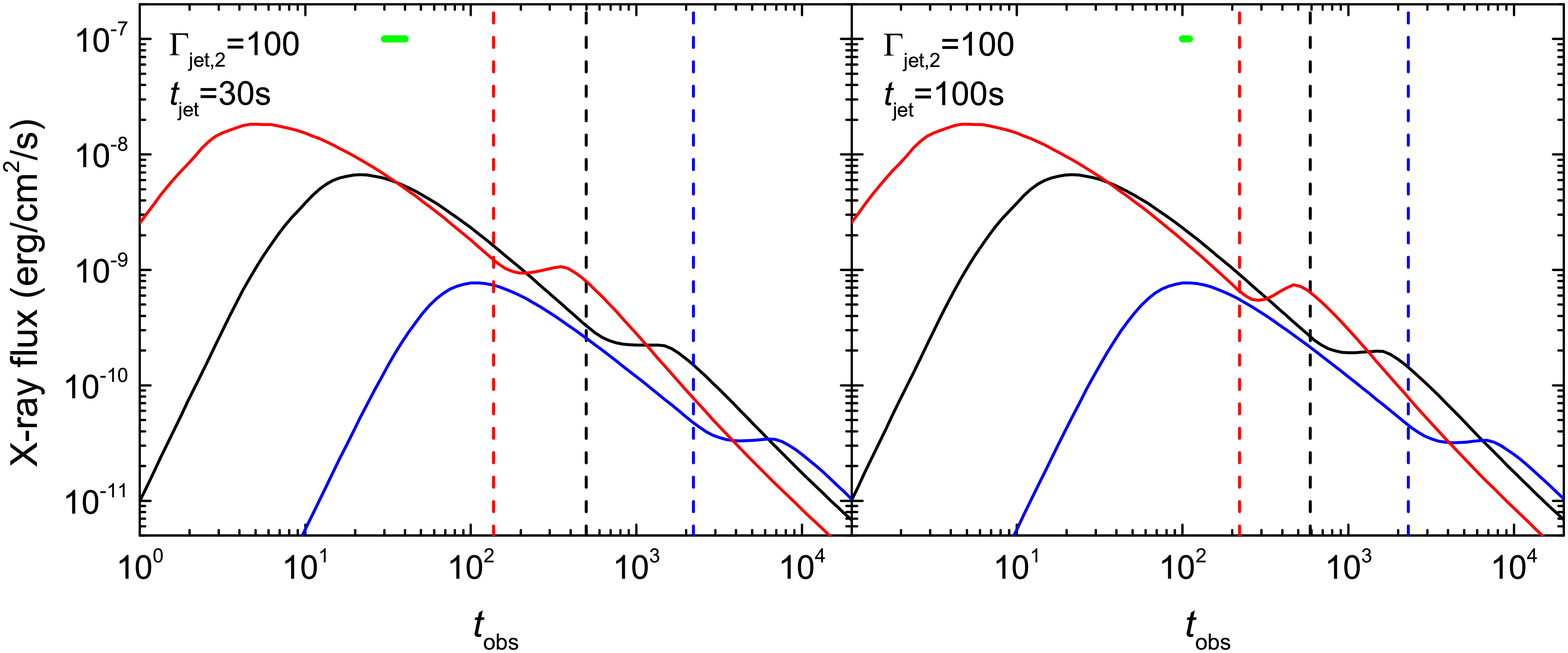}
\caption{Evolution of the external shock with energy injection
from the remnants of JET2,
where $t_{\rm jet}=30$~s ($100$~s) is assumed in the left (right) panels
and the value of $\Gamma_{\rm jet2}=300$ ($100$) is taken in the upper (lower) panels.
The horizontal green thick lines show the observed time of the second launched ejecta,
and the red, black, and blue solid lines (vertical dashed lines) represent the 0.3-10~keV X-ray flux ($t_b$)
in the situation with $n_0=10^2$, $1$, and $10^{-2}$, respectively.
}\label{Fig:fig1}
\end{figure}

\begin{figure}
\plotone{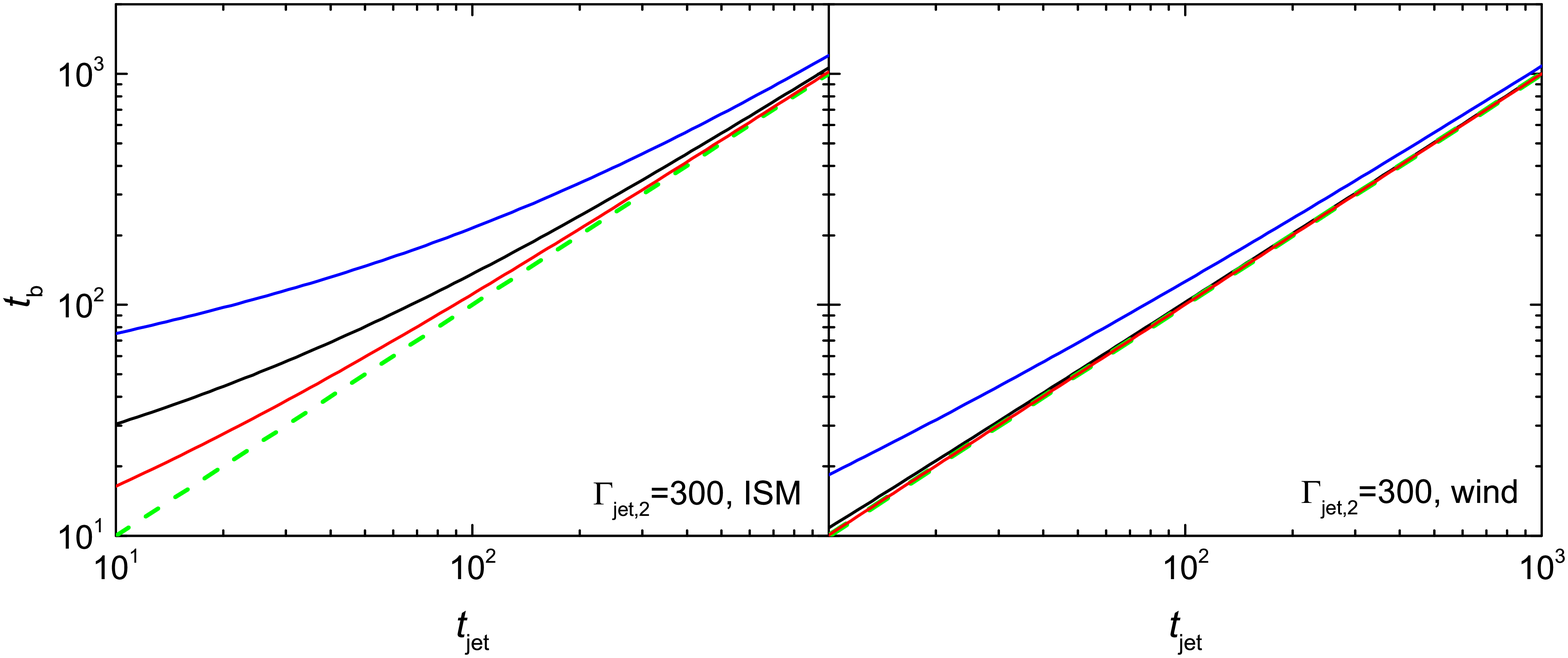}
\plotone{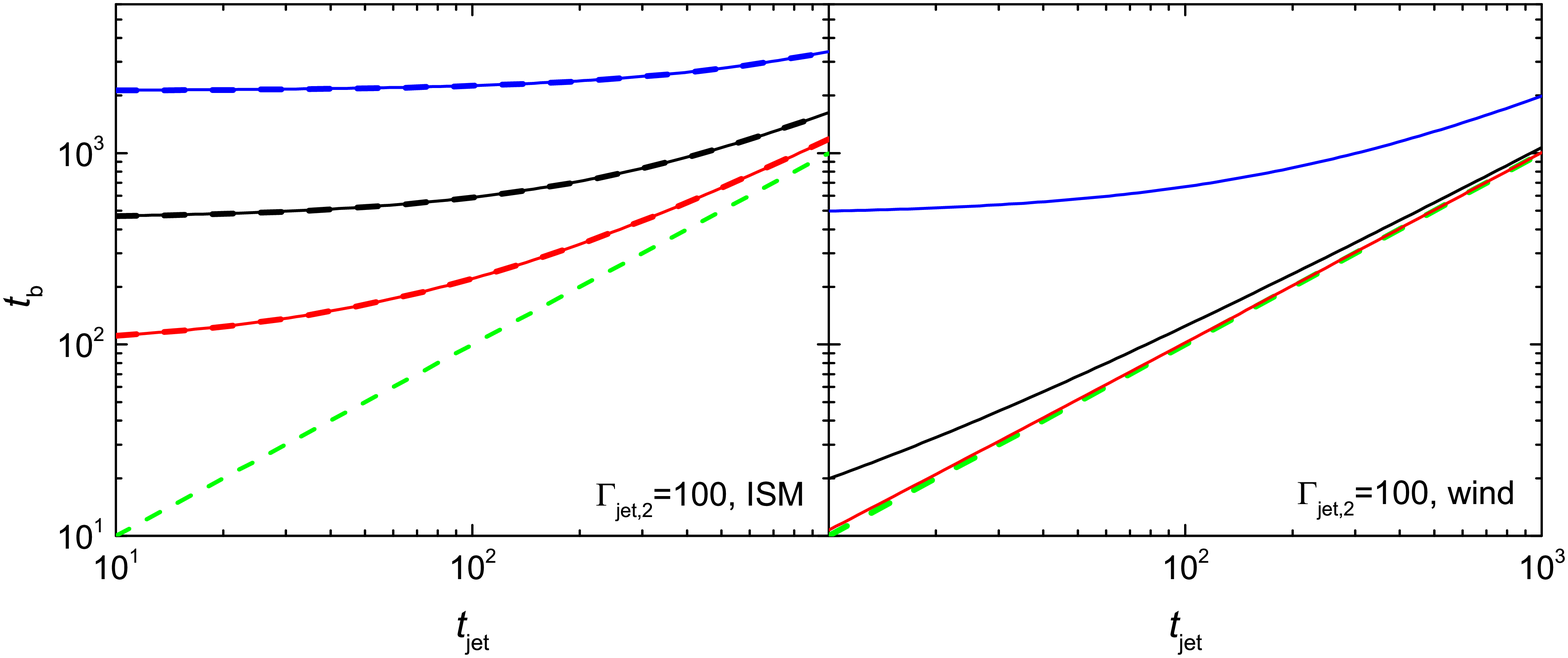}
\plotone{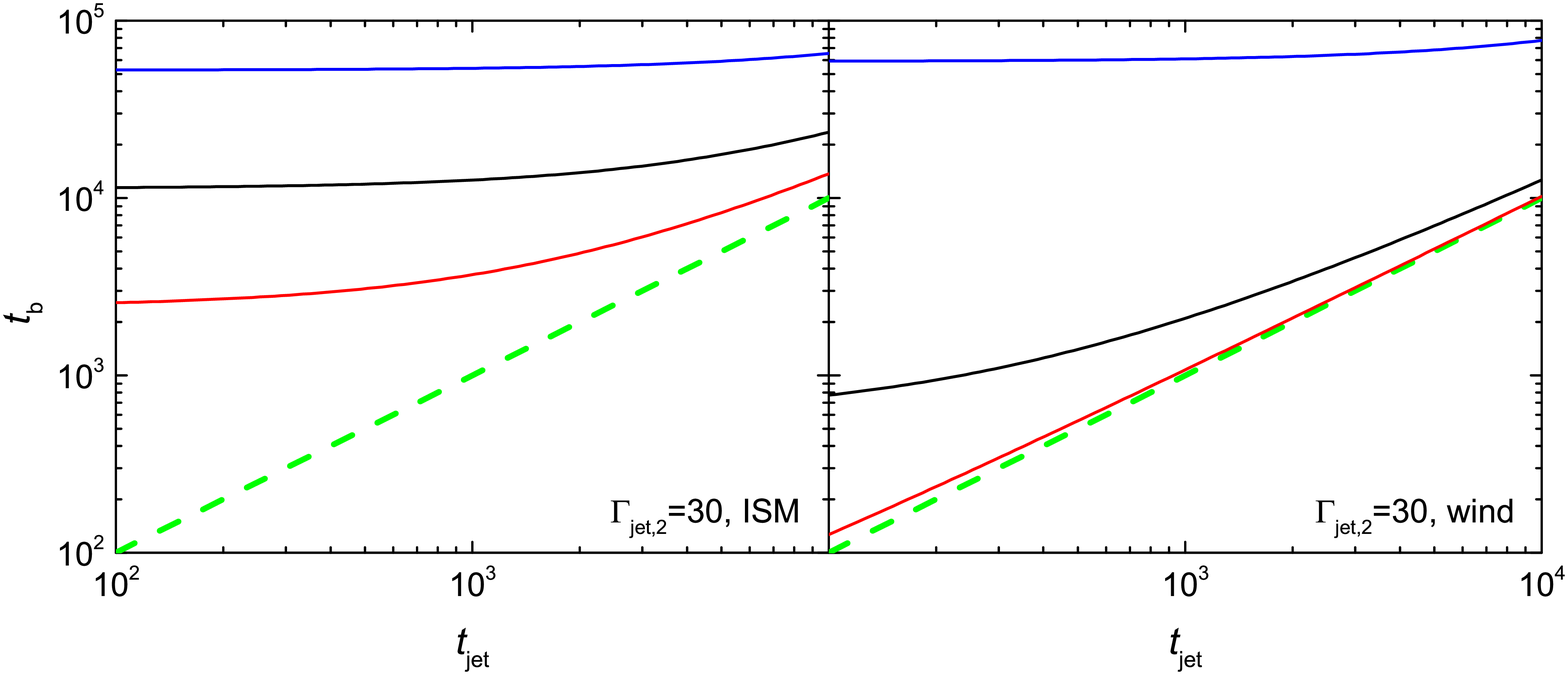}
\caption{Comparison of $t_b$ and $t_{\rm jet}$.
Here, the green dashed lines describe the relation of $t_b=t_{\rm jet}$,
the left (right) panels describe the situations with ISM (wind) environment,
and
$\Gamma_{\rm jet2}=300$, $100$, and $30$ are adopted in the upper, middle, and lower sub-figures, respectively.
The blue, black, and red solid lines in the left (right) panels
represent the situations with $n_0\,(A_*)=10^{-2}$, $1$, and $10^{2}$, respectively.
}\label{Fig:fig2}
\end{figure}

\begin{figure}
\plotone{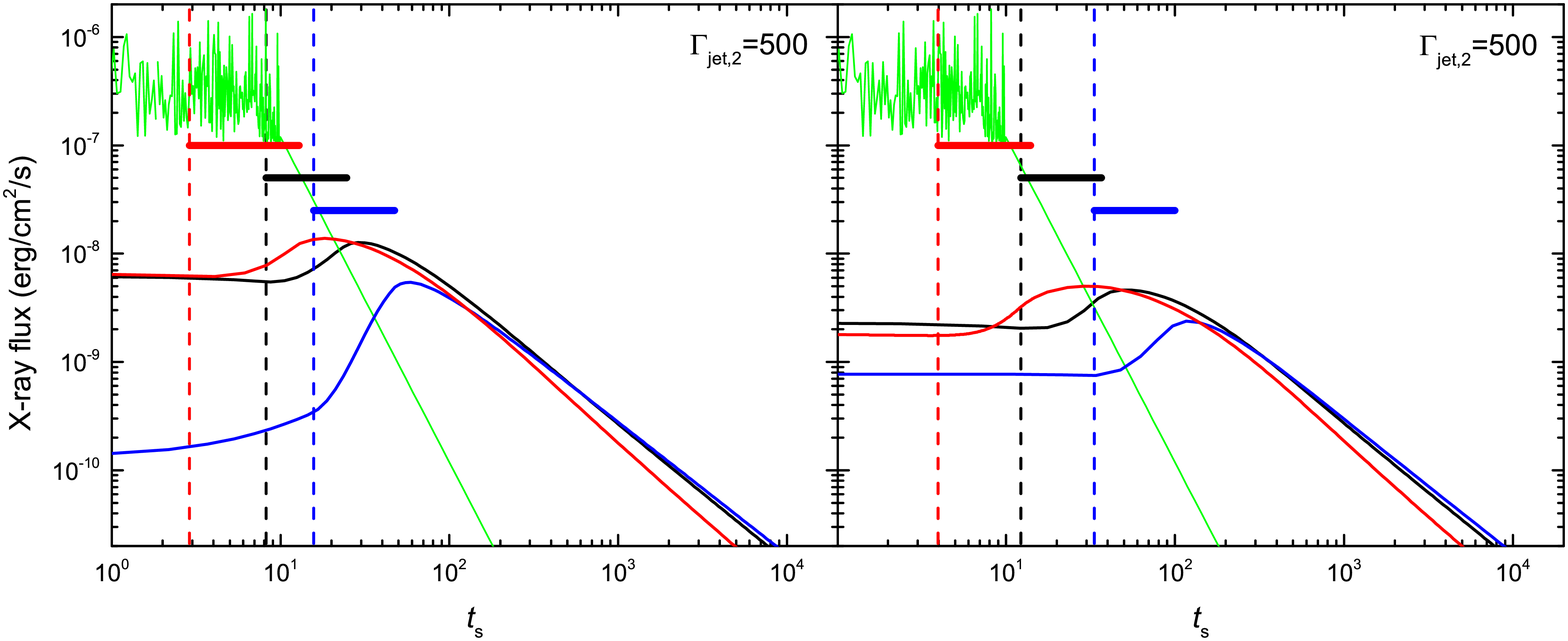}
\plotone{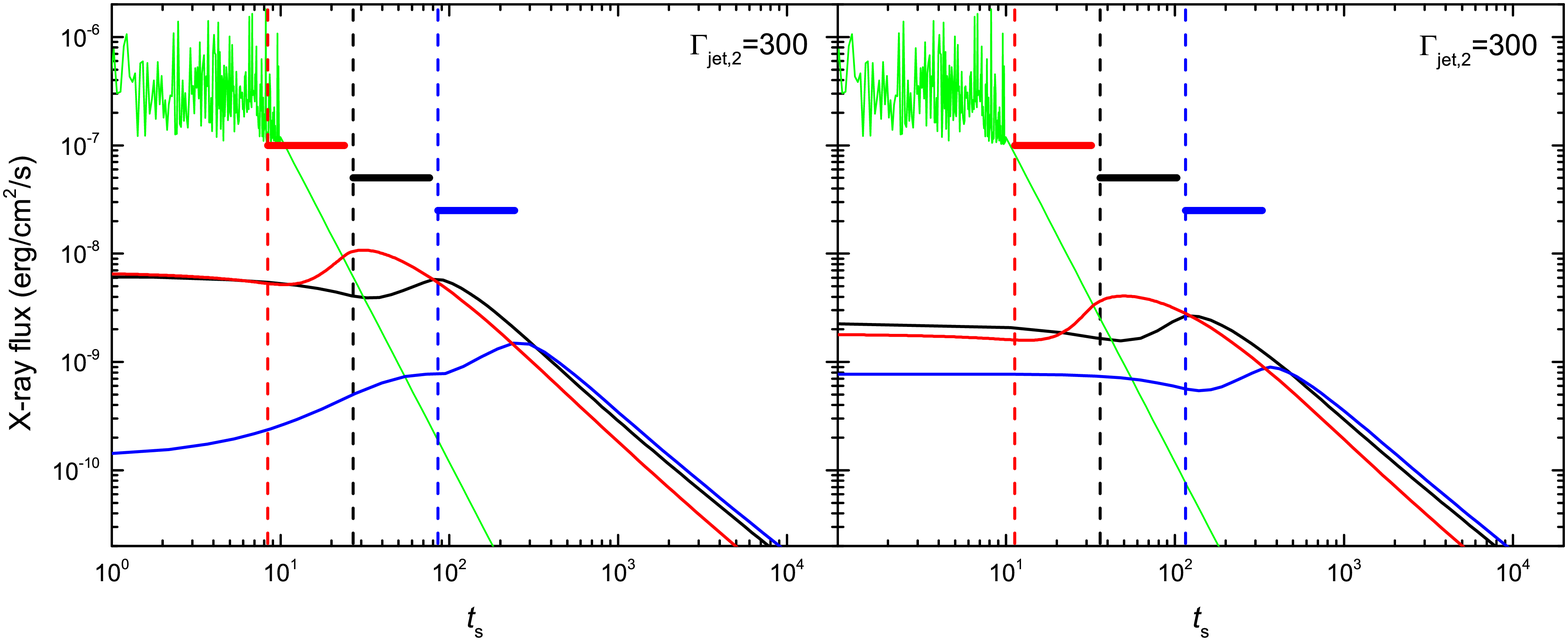}
\plotone{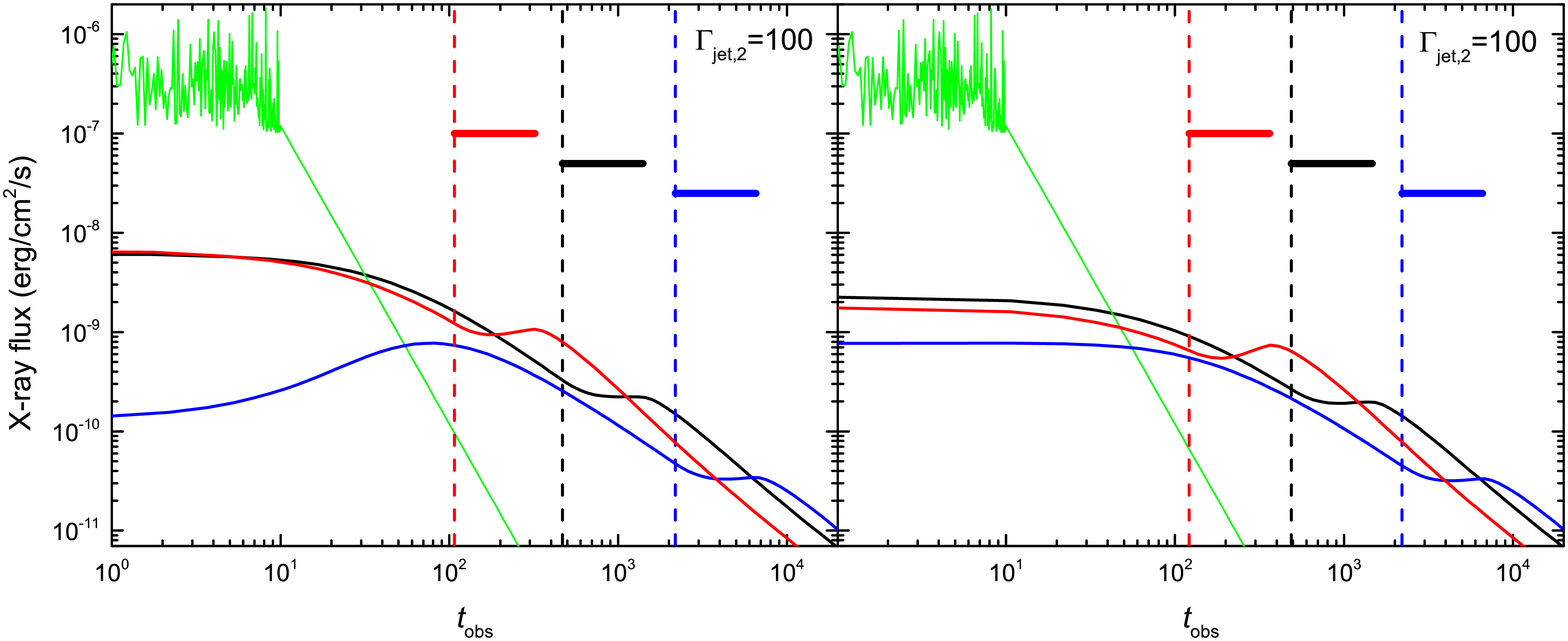}
\caption{Synthetic X-ray light curve from the burst trigger time for GRBs with a precursor,
where the main prompt emission (green solid lines) triggers the observation of BAT and
the precursor is assumed to be observed at $t_{\rm obs}=-30$~s ($-100$~s) in the left (right) panels.
The thick red, black, and blue horizontal lines indicate the energy injection phase of the external shock
and the meanings of other lines are the same as those in Figure~\ref{Fig:fig1}.
}\label{Fig:fig3}
\end{figure}

Recently, an extremely bright GRB~160625B was detected by Fermi Gamma-Ray Burst Monitor and Large Area Telescope.
Its prompt $\gamma$-ray lightcurve is composed of three episodes: a short precursor,
a very bright main emission episode, and a weak later emission episode (\citealp{Zhang_BB-2016-Zhang_B}).
The three episodes emission are separated by two long quiescent intervals.
Since the released energy in the first episode is lower than that in the second episode (\citealp{Zhang_BB-2016-Zhang_B}),
the value of $t_{\rm b}$ would be at around $t_{\rm jet}$.
Here, we assume that the bulk Lorentz factor of ejecta is proportional to the observed isotropic energy (e.g., \citealp{Lu_HJ-2014-Zhang_B-Liang}; \citealp{Liang_EW-2015-Lin_TT}).
That is to say, the remnants of the JET2
colliding with the external shock
and the corresponding radiative signature would appear at around the observed time of the main prompt emission.
An optical flash formed in this collision is indeed found
in the main prompt emission phase (\citealp{Lu_HJ-2017-Lu_J}).
The situation is also applicable for GRB~140512A (\citealp{Huang_XL-2016-Xin_LP}) and other bursts with a precursor.
We note that the \emph{Swift}/BAT is only triggered at the main prompt emission phase for some GRBs with a precursor (e.g., \citealp{Hu_YD-2014-Liang_EW}).
Then, we show the synthetic X-ray light curve from the burst trigger time for these GRBs.
The results are shown in Figure~\ref{Fig:fig3},
where the main prompt emission with isotropic energy $E_{\gamma,\rm iso}=10^{53}{\rm erg}$ is shown with green solid lines,
and $\Gamma_{\rm jet2}=500$, 300, and 100 are adopted in the upper, middle, and lower panels, respectively.
In this figure, the horizontal thick solid lines indicate the phase of the energy injection
and the meanings of other lines are the same as those in Figure~\ref{Fig:fig1}.
In addition, the precursor is assumed to be observed at $t_{\rm obs}=-30$~s ($-100$~s) in the left (right) panels of Figure~\ref{Fig:fig3}.
One can find that the light curve of X-rays does not behave as the form of a single bump.
The energy injection into the external shock
and thus an RS may appear in the phase of main prompt emission.
Then, more than one RS may appear in or after the main prompt emission phase,
e.g., GRB~130427A (\citealp{Vestrand_WT-2014-Wren_JA}).
One should be careful in dealing with the RS, especially for that in GRBs with a precursor.
It is valuable to point out that the RS
may also appear during the propagation of the JET1 into the circum-burst medium.
In this situation, the evolution of the RS is discussed in two cases: thick- and thin-shell cases
(e.g., \citealp{Sari_R-1995-Piran_T,Wu_XF-2003-Dai_ZG,Zou_YC-2005-Wu_XF,Granot_J-2012,Yi_SX-2013-Wu_XF,Fraija_N-2015}).
In the thick-shell case, the RS becomes relativistic during its propagation and the JET1 is significantly decelerated.
The emission of the RS is overlapped with the emission of JET1.
In the thin-shell case, the RS cannot decelerate the JET1 effectively
and thus the emission of the RS may be lagged behind the emission of JET1.
However, there is no energy injection (into the external shock) associated with the RS in this situation.

If the circum-burst environment and the energy injection time have been estimated,
one can estimate the Lorentz factor of the ejecta producing X-ray flare/plateau/bump,
and even those producing the main prompt emission for GRBs with a precursor.
In Figure~\ref{Fig:fig4},
we estimate the relations of $t_{\rm b}$ and $\Gamma_{\rm jet2}$ for GRB~160625B in different situations,
where $t_{\rm b}$ is the observed time of JET2 (i.e., the jet producing main prompt emission) colliding with the formed external shock and $z=1.406$ (\citealp{Xu_D-2016-Malesani_D}) is adopted.
In this figure,
the situations with an ISM environment and $n_0=36$ (\citealp{Lu_HJ-2017-Lu_J})
are shown with black lines,
the situations with a wind environment and $A_*=0.2$ (\citealp{Fraija_N-2017-Veres_P})
are shown with red lines,
and the situations with $E_{\rm k,iso}=8.8\times 10^{52}$,
$8.8\times 10^{51}$, and $8.8\times 10^{50}$~erg
are plotted with the dashed, solid, and dash-dotted lines, respectively.
In GRB~160625B, an optical flash formed in the RS is found at the observer time $200$~s (\citealp{Lu_HJ-2017-Lu_J}).
Then, we would like to believe that the energy injection into the formed external shock from JET2 appears at
the observer time $t_{\rm b}=200$~s,
which is shown with the blue dotted line in Figure~\ref{Fig:fig4}.
It can be found that the value of $\Gamma_{\rm jet2}=220$ (107) are required for the situations with $E_{\rm k,iso}=8.8\times 10^{51}$~erg
and ISM (wind) environment.
It is worth noting another energetic burst GRB~130427A,
which was observed in GeV-MeV $\gamma$-rays, X-rays, and the optical band.
In order to explain the multiwavelength observations,
\cite{Vestrand_WT-2014-Wren_JA} claimed that more than one episode of energy injection and RS were necessary.
In the time interval from 9.31 s to 19.31 s after the GBM trigger,
a bright optical flash with a magnitude of $7.03\pm 0.03$
is reported by RAPTOR (\citealp{Vestrand_WT-2014-Wren_JA}).
In addition, a bright LAT peak in coincidence with this optical flash
is found (\citealp{Ackermann_M-2014-Ajello_M}).
Then, \cite{Fraija_N-2016-Lee_W} interpreted the bright optical flash/the extreme LAT peak
as the synchrotron/synchrotron self-Compton emission from the RS
when the deceleration of the ejecta evolves as that in the thick-shell case.
For the other RS in this burst, the energy injection into the external shock is required (\citealp{Vestrand_WT-2014-Wren_JA}).
Since the jets being responsible for the energy injection are not observed definitely
(see Figs. 1 and 2 in \citealp{Ackermann_M-2014-Ajello_M}),
the relation of $t_b$ and $t_{\rm jet}$ could not be estimated based on the observations of the other RS.

\acknowledgments
We thank Bing Zhang for helpful discussions.
This work is supported by
the National Basic Research Program of China
(973 Program, grant No. 2014CB845800),
the National Natural Science Foundation of China (grant Nos. 11773007, 11403005, 11533003, 11573023, 11473022),
the Guangxi Science Foundation (grant Nos. 2016GXNSFDA380027),
the Special Funding for Guangxi Distinguished Professors (Bagui Yingcai \& Bagui Xuezhe),
and
the Innovation Team and Outstanding Scholar Program in Guangxi Colleges.


\clearpage

\clearpage

\clearpage

\clearpage
\begin{figure}
\plotone{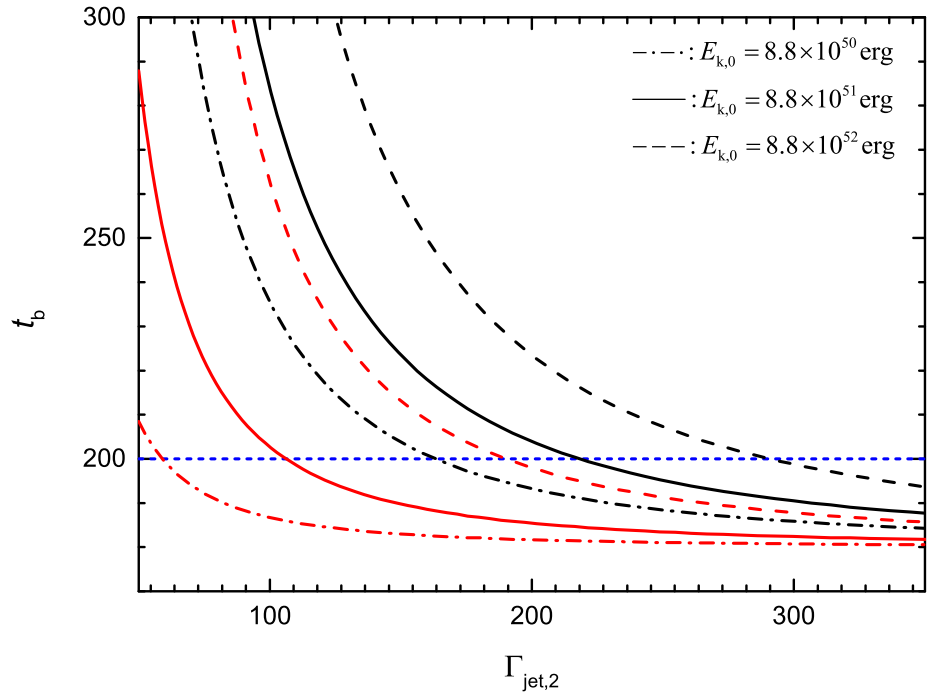}
\caption{Relation of $t_{\rm b}$ and $\Gamma_{\rm jet2}$ for GRB~160625B,
where $t_{\rm b}$ is the observed time of JET2 hitting the external shock.
The black lines are for the situations with an ISM environment and $n_0=36$,
the red lines are for the situations with a wind environment and $A_*=0.2$,
and the dotted blue line indicates the observed time of the optical flash, i.e., $t_{\rm b}=200$~s.
}
\end{figure}\label{Fig:fig4}

\end{document}